\documentstyle[twoside,fleqn,espcrc2,epsf,rotate]{article}

  \newcommand{\cZ}{\cal{Z}}

\newcommand{\AmS}{{\protect\the\textfont2
  A\kern-.1667em\lower.5ex\hbox{M}\kern-.125emS}}

\title{RESULTS ON FINITE DENSITY QCD}

\author{ Ian M. Barbour, Susan E. Morrison, 
\address{Dept. of Physics and Astronomy, University of Glasgow,
\\ G12 8QQ, U.K.\\ UKQCD Collaboration}
Elyakum G.~Klepfish,
\address{Dept. of Physics, King's College London, London WC2R 2LS}
John B. Kogut  
\address{Department of Physics, University of Illinois, 1110 West Green Street,\\
Urbana, IL 61801}
and
Maria-Paola~Lombardo
\address {Zentrum f\"ur interdisziplin\"are Forschung,
  Universit\"at Bielefeld,  D-33615 Bielefeld, Germany}}

\begin{document}

\begin{abstract}
A brief summary of the formulation of QCD at finite chemical
potental, $\mu$, is presented. The failure of the quenched approximation to the problem
is reviewed. 

Results are presented for dynamical simulations of the theory
at strong and intermediate couplings. We find that the problems associated with
the quenched theory persist: the onset of non-zero quark number does seem
to occur at a chemical potential $\approx { {m_{\pi}} \over 2}$. However analysis of the Lee-Yang zeros of
the grand canonical partition function 
in the complex fugacity plane, ($e^{\mu/T}$), does show signals of critical
behaviour in the expected region of chemical potential.

Results are presented for a simulation at finite
density of the Gross-Neveu model on a $16^3$ lattice near to the chiral limit. 
Contrary to our simulations of QCD no pathologies
were found when $\mu$ passed through the value ${ {m_{\pi}}\over 2}$.
\end{abstract}

\maketitle

\section{Introduction}
   Lattice QCD at finite baryon density holds the key to an understanding of
the phase transition between quark-confined hadronic matter 
and the quark-gluon plasma. 
At zero temperature we expect this 
transition to occur at $\mu = { {m_p}\over 3}$ 
corresponding to the lowest lying state with
non-zero baryon number. \\
Although the QCD phase transition at finite temperature has been
successfully studied on the lattice with simulations predicting a first 
order transition at a well defined critical temperature,
a non-perturbative lattice calculation of nuclear matter at finite
density has been an outstanding problem in QCD thermodynamics since
1986\cite{barbourstone86,KLS95}.	The fundamental difficulty in simulating QCD at
finite density and investigating the transition quantitatively is that
the effective action resulting from the Grassmann integration over the
fermions is complex due to the introduction of the chemical
potential in the Dirac matrix. This complex nature of the QCD finite
density action\cite{kogmatstone83,hazkar83} prohibits the use of naive 
probabilistic methods in evaluating the functional integral. Thus the
standard Hybrid Monte Carlo simulation algorithms
for lattice QCD with dynamical fermions are inappropriate in this
context since they require a positive definite probability measure.

The study of lattice QCD at finite density is motivated by the need for
a full equation of state for nuclear matter in the temperature-chemical
potential plane which would give quantitative predictions for critical
energies and densities in relativistic heavy ion collisions. Other
applications include equations of state for neutron stars and
cosmological models of the early universe.

\section{Quenched Simulations}
Serious problems were first reported in quenched simulations of finite density
QCD in 1986\cite{barbourstone86} and the physical and mathematical reasons for
this failure have been the focus of considerable debate ever since
\cite{KLS95,KLS96}.

	In these early simulations  the behaviour of the chiral condensate was studied
at fixed quark mass for various different values of 
the chemical potential, $\mu$. The behaviour
was initially as expected {\it viz.} the chiral condensate remained constant up until a
certain value of mu and then tended to zero as the chemical potential was
increased.

 We would expect that physical observables are $\mu$ independent up to some ${\mu}_c$
which is related to the threshold for baryon production. The problem encountered was
the following: as the quark mass was decreased the 
$\mu$ at which  the chiral condensate began to change also decreased towards
zero.

 In fact, the onset of chiral symmetry restoration appeared to occur at a chemical potential
of half the pion mass, which would extrapolate to zero in the chiral
limit. A study of the distribution of the eigenvalues of the lattice
Dirac operator confirmed that the lowest mass state containing a net
number of fermions (i.e. a quark or a baryon but not a meson which does
not see $\mu$) became massless as the bare quark mass was reduced to zero.
The interpretation was that there existed either baryonic states which became massless
in the chiral limit and had an energy equal to ${\frac{3}{2}}m_{\pi}$ or stable quark matter
with low mass per baryon.

	The result that ${\mu}_c$ is proportional to the pion mass is clearly
unphysical.We expect ${\mu}_c \simeq { {m_p}\over 3}$ because the proton
is the lightest state with non-zero baryon number. From this we are drawn to two
possible conclusions. The first possibility is that the quenched approximation is at fault
so that it is strictly necessary  to consider the complex action of full QCD at finite
chemical potential. Secondly, there could be  intrinsic problems in the lattice formulation of
fermions (possibly associated with fermion doubling) and chemical potential which would
survive an unquenched treatment.

Further studies\cite{DAVKLEP91} of the
quenched theory on larger lattices found similar behaviour. However, a recent study
\cite{KLS95,KLS96} of the quenched theory which measured the condensate and the pion
and nucleon masses, did find that, at intermediate and strong coupling, the theory
is sensitive to the baryon mass but that it is pathological for $\mu > {{m_{\pi}}\over 2}$.

From an analytical study of the eigenvalues of the fermionic propagator matrix
Gibbs\cite{GIBBS86} concluded that the eigenmodes of the  propagator
matrix (defined below) relate to the mass spectrum of the theory. The Gibbs argument,
requires the calculation of the hadronic spectrum on replicated
lattices, i.e. lattices, with periodic boundary conditions on the 
gauge fields, which have been strung together $d$ times in
the time direction. He considered the limit $d \to \infty$ in order to
replace finite sums with contour integrations.
The procedure is justifiable at zero temperature. The expression obtained for
the inverse of the duplicated fermion matrix $G(t_1,t_2)$ is

\begin{equation}
G(t_1,t_2) = \sum A_a \lambda_a ^{t_1 - t_2}
\end{equation}
where $A_a$ are the amplitudes (which can be related to the
eigenvectors of the propagator matrix) and $\lambda_a$ are the associated eigenvalues. 
 The form of the inverse shows that the exponential decay at large separation is
controlled by the eigenvalues $\lambda_a$ : thus we see
that the eigenvalue spectrum calculated on isolated configurations
should contain poles in correspondence to the physical masses.
In particular, the smallest mass state can clearly be extracted from the
lowest eigenvalues. This state is obtained by the squaring  the propagator
matrix, and defines the pion mass in QCD. This implies that in the
quenched theory $<\bar{\psi}{\psi}>$ is controlled by the {\it pion mass}.

It has also been suggested by several authors that the coincidence of the onset of
the chiral symmetry restoration with one half of the pion mass might only have been a
numerical accident, the correct relationship being 
$ {\mu}_{onset}={{m_{N}}\over {3}} - \Delta$ where $\Delta$ is the contribution of
the nuclear binding energy. If this scenario were true then the problems with finite
density simulations would not be too serious. This conjecture has been tested 
\cite{KLS96}. However this work confirmed the onset at $\mu= { {m_{\pi}}\over
2}$.

Recent work by Stephanov \cite{Steph96} using the
random matrix method has 
provided some insight into this particular problem by demonstrating that at nonzero $\mu$ the
quenched theory is not a simple $n_f \rightarrow 0$ limit of a theory with $n_f$
quarks but it is the $n_f \rightarrow 0$ limit of a theory with $n_f$ quarks
and $n_f$ conjugate quarks and therefore inappropriate to QCD.This
explanation implies that the phase of the determinant is important for a simulation of full QCD
at finite $\mu$. In the quenched model the early onset for the number density
would also correspond to the restoration of chiral symmetry because of
the simultaneous occurrence of quarks and conjugate quarks in the system
whereas the inclusion of dynamical fermions would result in a
rearrangement of the eigenvalues such that the chiral symmetry would be
restored at $\mu_c\simeq { {m_{B}}\over 3}$ as expected.

\section{Methods for Simulating QCD at $\mu \neq 0$ with Dynamical
Fermions}
The pathologies of quenched studies of the chiral phase transition at
finite density indicate that a correct implementation of dynamical 
quarks in the simulations is essential to the physics of the model.

\subsection{The Glasgow Method} 
The problem caused by the non-Hermitian nature of the
fermion matrix at $\mu \neq 0$ can be circumvented by a method which
involves the expansion of the grand-canonical partition function (GCPF)
in powers of the fugacity variable $e^{\mu/T}$.

The GCPF can be expressed as an ensemble average of the fermionic determinant
normalised with respect to the fermionic determinant at $\mu=0$.  

\begin{equation}
{\cZ}={{\int [dU][dU^\dagger]|M(\mu)|e^{-S_g[U,U^\dagger]}}\over {
\int [dU][dU^\dagger]|M(\mu=0)|e^{-S_g[U,U^\dagger]}}}
\end{equation}
where $S_g$ is the pure
gauge action and $M$ is  the Kogut-Susskind fermion matrix.

This means that the probability measure is well-defined and standard Hybrid
Monte-Carlo algorithms can be applied. The GCPF is then given by the
ensemble average

\begin{equation}
{\cZ}=\left<{{|M(\mu)|}\over {|M(\mu=0)|}}\right>_{\mu=0}
\end{equation}

The Kogut-Susskind fermion matrix has the structure:

\begin{equation}
2iM=G+ 2im + \hat V e^{\mu}+\hat V^{\dagger} e^{-\mu}.
\end{equation}
in which $G$ contains all the spacelike gauge links and the bare 
quark mass and $\hat V$ ($\hat V^{\dagger}$) all the
forward (backward) timelike links.

It then follows that, on a lattice of size $n_s^3n_t$, the determinant 
of $M$ is given by\cite{GIBBS86}

\begin{equation}
|2iM| = e^{3\mu n_s^3 n_t} |P - e^{-\mu}|
\end{equation}
where the propagator matrix $P$ is given by

\begin{equation}
P=\left(\begin{array}{cc}
-(G+2im)\hat V & \hat V \\
    -\hat V    & 0
\end{array} \right)
\end{equation}
with its inverse given by

\begin{equation}
P^{-1}=\left(\begin{array}{cc}
0           & -\hat V^{\dagger}\\
\hat V^{\dagger} & -\hat V^{\dagger}(G+2im)
\end{array} \right)
\end{equation}

From the structure of the propagator matrix and its inverse, it is easily shown that its
eigenvalues have the symmetry that if $\lambda$ is an eigenvalue then so is
$1/\lambda^\dagger$. In addition, since the matrix $\hat V$ factors, the eigenvalues have a
$Z(n_t)$ symmetry. As a consequence of this symmetry, the characteristic polynomial
for $P$ is a polynomial in $e^{\mu n_t}$ with $6n_s^3+1$ complex coefficients. Since
the probability for a given configuration of gauge fields to appear in our ensemble is
equal to that of its complex conjugate their imaginary parts will average to zero. We
impose this symmetry in the simulations described below. As a consequence, the
characteristic polynomial from a configuration (averaged with its complex conjugate) is
invariant under $\mu \rightarrow -\mu$ and has the form:

\begin{equation}
   \sum_{n=-3{n_s}^3 }^{3{n_s}^3} b_{|n|} e^{n\mu {n_t}}
\end{equation}

The chemical potential dependence of the GCPF is then given explicitly by
\begin{eqnarray}
{\cZ}&=&\sum_{n=-3{n_s}^3 }^{3{n_s}^3}\left<b_{|n|}\right>e^{n\mu {n_t}} \nonumber \\
&=&\sum_{n=-3{n_s}^3}^{3{n_s}^3}e^{-(\epsilon_n - n\mu)/T}
\end{eqnarray}
with $e^{-\epsilon_n n_t}=<b_{|n|}>$

By measuring the $\epsilon_n$ in the simulations we can readily obtain the fermion number
density, \\ 
$\rho=<n>$ and the energy density, $E=<\epsilon_n>$.

	The relative values of the canonical partition functions can be
used to characterize the properties of the system. In particular the
relative value of the triality non-zero coefficients give an indication
of whether the system is in the confined or deconfined regime. In the
confined sector the ensemble average of the triality non-zero
coefficients should tend to zero as the statistics increase. We see strong evidence
for this behaviour is seen in the lattice QCD simulations described below. 

	 We also consider a description of the system in terms of  the canonical partition 
functions for fixed particle number. This technique allows us to
include only those triality zero coefficients which, at the end of the simulation,
are positive and have a reasonable error.The chemical potential as a function of the
baryon number density 
is obtained from the local derivative 
of the energy, $\epsilon_n$, of the state with $n$ fermions with respect to $n$. 
\begin {equation}
\mu(\rho) = {{1}\over {3 n_s^3 }}{{\partial \epsilon_n}\over{\partial\rho}}
\end {equation}
where $\rho$ is the fermion density, $n\over{3n_s^3}$.
	
\subsection{Lee-Yang Zeros}
	According to the theorems of Lee and Yang\cite{LY} the phase structure of
a simulated system can be determined by studying the zeros of the GCPF.
The zeros correspond to singularities in the thermodynamic potential.
We study the zeros of the GCPF in the complex chemical potential (or
fugacity plane) which are given by $Z(\mu)=0$. According to the Lee-Yang
theorem as the system volume tends to infinity, the zeros will converge
towards the critical value in the physical domain (real axis) of the
chemical potential.

\subsection{Alternative Methods}

	The Glasgow method involves an ensemble generated at $\mu=0$.
This may introduce systematic errors due to insufficient overlap when
measuring observables at $\mu$ significantly greater than the onset of
non-zero number density. An alternative method involves generating the
ensemble either with respect to the absolute value of $ |M(\mu)| $
or with respect to the modulus of its real part\cite{Vladikas88,Toussaint90}.
In these simulations it is necessary to measure an additional observable
related to the average sign of the determinant. For example, updating with respect to
the modulus gives the observable ${\cal{O}}$ as

\begin{equation}
{\cal O}=\frac{
{\left<{\cal O}e^{i\phi}\right>}_{|M|}
}{{\left<e^{i\phi}\right>}_{|M|} }
\end{equation}
where $\phi$ is the phase of the determinant. Simulations at a fixed quark
mass using these methods are well behaved at low and high chemical potentials
but for ${ {m_{\pi}}\over 2} \le \mu \le \approx {{m_p}\over 2}$ the phase
$\phi$ fluctuates violently from configuration to configuration and $<e^{i\phi}>$
becomes very small and is not measurable. This is the region of $\mu$ in which
critical behaviour is expected.

Gocksh\cite{Gocksh88} used a spectral density method by binning the phase 
$\phi$ and measuring $\rho(\phi)$. On a $2^4$ lattice, his results for $\mu_c$
are in broad agreement with meanfield.

\section{The Static Quark Model}

Blum, Hetrick and Toussaint\cite{BHTous96} have recently studied numerical
simulations of lattice QCD in the limit that the quark mass and chemical potential
are simultaneously made large. In that limit, the quark mass and chemical potential
appear only in the ratio $(2ma/e^{\mu a})^{n_t}$ and the propagator matrix becomes 
\begin{equation}
P=\left(\begin{array}{cc}
-2im\hat V & \hat V \\
  -\hat V  & 0
\end{array} \right).
\end{equation}
The corresponding fermion determinant is complex but
trivial to evaluate which allows generation of very high 
statistics in their measurements and determine $< e^{i\phi}>$ to sufficient accuracy.

For quenched QCD the high temperature transition is first order and they 
expected this behaviour to extend into the interior of the $T-\mu$ phase diagram.
However their simulations showed that this transition becomes a smooth 
crossover at very small density (possibly for any nonzero density) and that,
at low enough temperature, chiral symmetry remains broken at all densities.

Of course, as the authors point out,it is not at all clear that the static approximation has anything to
do with real QCD. However, it is relevant that unexpected results follow in this
simple model.

\section{Lattice QCD at Strong Coupling}

	The strong coupling limit of QCD provides us with an important
testing ground for lattice Monte-Carlo simulations at finite density because
we can compare our results with the analytic predictions of 
mean-field theory. No such analytic predictions are available in the
scaling region (i.e. intermediate coupling). At infinite coupling the 
mean-field method predicts a {\it first order} transition at chemical
potential $\mu_{mf}$ where

\begin{equation}
\mu_{mf}={1\over{r}}\sinh^{-1}(\lambda_{0}r)-{{{\lambda_{0}}^2}\over
{(d-1)r}}
\end{equation}
where 
\begin{equation}
\lambda_{0}={{1}\over{r\sqrt2}}{\left(\sqrt{1+{(d-1)}^{2}r^{4}} -1\right)}^{1/2}
\end{equation}
for a lattice with $d$ space-time dimensions where $r=n_t/n_s$.

The mean-field baryon and pion masses are given by

\begin{equation}
 m_B = \ln\left[{1\over2}c^3+\sqrt{1+{1\over8}c^6}\right]
\end{equation}

\begin{eqnarray}
 m_{\pi} & = & \ln\left[1+{1\over2}(c^2-2d)   \right.\\   
         &   &  \left.+\sqrt{(c^2-2d)+{1\over4}(c^2-2d)^2} \right]
\end{eqnarray}
where
\begin{equation}
c=(m+ \sqrt{2d+m^2})
\end {equation}
Note that  $\mu_{mf}$ does not correspond to the mean-field baryon threshold $m_B/3$.	

 It has been argued that the discrepancy is
related to the binding energy of nuclear matter which is large at
infinite coupling. Bilic et al.\cite{Bilic92} demonstrated explicitly that $1/g^2$
corrections diminish the discrepancy between  $\mu_{mf}$ and the mass
of the lightest baryonic state divided by $N$ (the number of colours). The mean-field pion thresholds
and $\mu_{mf}$ are not well separated for bare quark masses $ma>0.5$.
 
	Monte-Carlo SU(3) simulations\cite{KarMut89} using the monomer-dimer algorithm
whereby the partition function is represented in terms of monomers,
dimers and baryonic loops estimate a critical chemical potential,
$\mu_c$ which is in agreement with  $\mu_{mf}$. This algorithm involves
integrating out the gauge fields exactly and subsequently integrating
the quark fields. Hybrid Monte-Carlo techniques integrate the quark
fields prior to the gauge fields. The monomer-dimer algorithm cannot be
used to explore small bare quark masses $m<0.1$ and is restricted to the
infinite coupling regime where the quarks are point-like.

 \subsection{Strong Coupling using the Glasgow Method} 

Consider again the expression for the GCPF

\begin{equation}
{\cZ}=\sum_{n=-3{n_s}^3}^{3{n_s}^3}e^{-(\epsilon_n - n\mu)/T}
\end{equation}
The zeros of this polynomial are the Lee-Yang zeros of $Z$ in the complex fugacity plane
(or equivalently the complex $e^{\mu}$ plane).
In the dynamical theory these zeros play a role analogous to that of the eigenvalues of
$P$ in the quenched theory when calculating the number density.

Z can be rewritten as

\begin{equation}
Z= e^{3n_s^3 \mu/T}\prod_{i=1,6{n_s}^{3}}(z-\alpha_i)
\end{equation}
where $z=e^{-\mu/T}$ and the $\alpha_i$ are the zeros of the {\it averaged}
 characteristic polynomial for $P$. 

On each configuration
\begin{equation}
\det{(M)}= e^{3n_s^3 \mu/T} \prod_{i=1,6{n_s}^{3}}(z-\lambda_{i})
\end{equation}
where the $\lambda_i$ are the eigenvalues of $P$.
Since ${Z}=\left<{{detM(\mu)}\over {detM(\mu=0)}}\right>_{\mu=0}$
we can interpret the zeros of the partition function as
the proper ensemble average of the eigenvalues of the fermionic
propagator matrix. 
One obvious constraint on the distribution of the zeros is
that in the confined sector
we expect to see a $Z(3)$ symmetry arising from  
the triality non-zero coefficients averaging to zero so that only canonical
partition functions containing multiples of three quarks contribute to
the thermodynamics. The $Z(3)$ symmetry constrains the phases of the
zeros but not their modulus. 

We calculate the fermion number density, $n_q$ from
\begin{equation}
<n_q> = {T \over V}  { {\partial \ln(Z)} \over { \partial \mu}}
\end{equation}

Hence in the quenched theory the number density, $n_q$, is given by
\begin{equation}
n_q = 1 + < { 1 \over {3V} } \sum_{-3V}^{3V} { {z} \over {\lambda_i-z }} >.
\end{equation}
where the $\lambda_i$ are the eigenvalues of $P^{n_t}$. 
Using the generalized boundary conditions arising from considering a configuration
on a replicated lattice one\cite{GIBBS86} can convert the sum to a contour integral
and obtain, for each configuration

\begin{equation}
n_d = 1/V \sum_{1 < |\lambda_i| < e^{\mu/T}} 1.
\end{equation}

In the dynamical theory the number density, $n_d$, is given by
\begin{equation}
n_d = 1 + { 1 \over {3V} } \sum_{-3V}^{3V} { {z} \over {\alpha_i-z }} 
\end{equation}
where the $\alpha_i$ are the zeros of the averaged characteristic polynomial for
$P$. Performing the contour integation as above gives
\begin{equation}
n_d = 1/V \sum_{1 < |\alpha_i| < e^{\mu/T}} 1.
\label{numbden}
\end{equation}
Thus if we consider plotting the zeros of the GCPF in the complex
$e^\mu$-plane, any discontinuities in the number density will be associated
with circular bands of increased density of zeros.

\subsection{Results at Strong Coupling}

We first consider, in the light of the above discussion concerning the
roles of the eigenvalues and the zeros of the GCPF, their distributions in
the complex $\mu$ and $e^{\mu}$ planes.

\begin{figure}[htb]
\epsfxsize=7.0cm
      \epsfbox{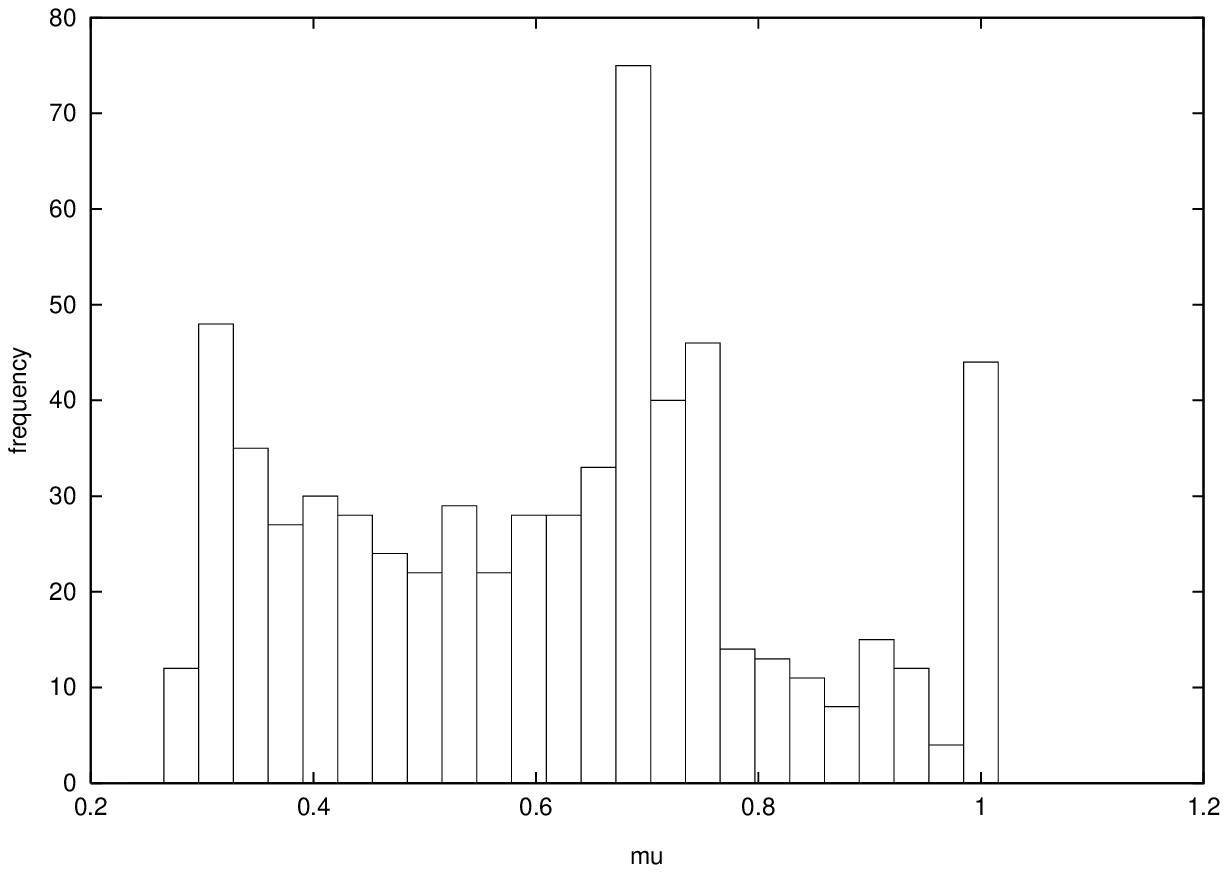}
\epsfxsize=7.0cm

       \epsfbox{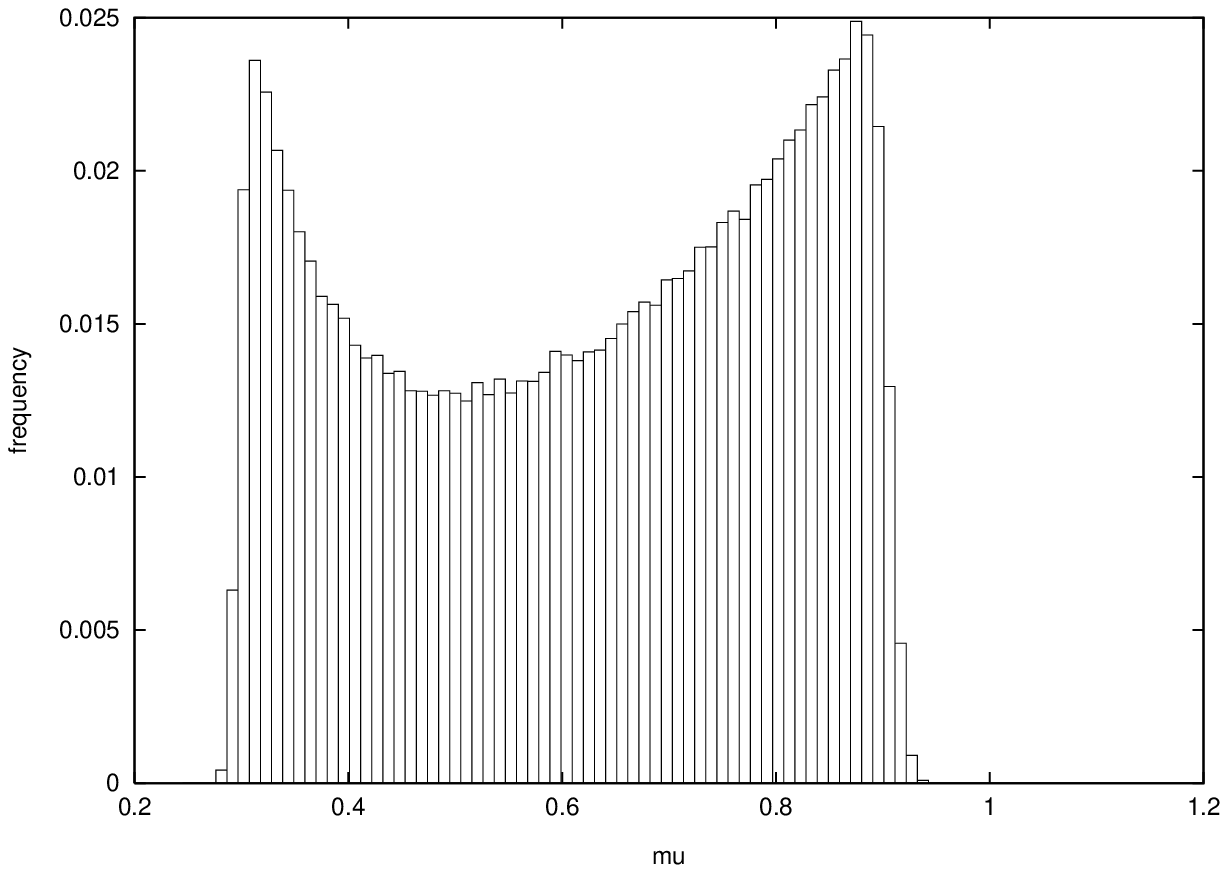}
    \caption[xxx]{ Histograms of the real parts of the eigenvalues and
of the real part of the zeros in the complex $\mu$-plane at strong coupling and bare
quark mass, $ma=0.08$ on a $6^4$ lattice.}
\label{fig:m0.08}
\end{figure}

Fig \ref{fig:m0.08} contains a histogram of the distribution of the positive real parts of the
eigenvalues in the complex $\mu$ plane. The distribution is obtained from 450
configurations on a $6^4$ lattice at a bare quark mass of 0.08. There is a 
clear signal that the number density becomes non-zero at $\mu \approx 0.3$ which
is consistent with $m_{\pi}/2$. This is the quenched onset $\mu$. The lattice
is filled at $\mu=0.95$.

The analogous distribution of the zeros of the GCPF is  also shown in Fig \ref{fig:m0.08}.
These zeros are those of the polynomial obtained by averaging 
the characteristic polynomials of the propagator
matrix over the same 450 configurations. There is still a clear signal that
the number density becomes non-zero at the same onset $\mu$ as that of the quenched
theory {\it but} a strong signal has developed via an intermediate peak in the distribution
which is absent from that of the eigenvalues. As argued above, this band of increased density
can be associated with a discontinuity in the number density.

\begin{figure}[htb]
\epsfxsize=7.0cm
\epsfbox{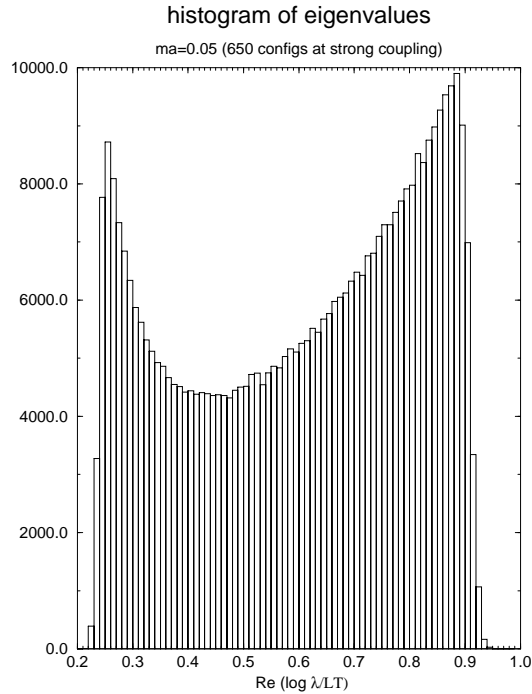}
    \caption[xxx]{ Histogram of the real parts of the eigenvalues 
in the complex $\mu$-plane at strong coupling and bare
quark mass, $ma=0.05$ on a $6^4$ lattice.}
\label{fig:eigm0.05}

\end{figure}

\begin{figure}[htb]
\epsfxsize=7.0cm
\epsfbox{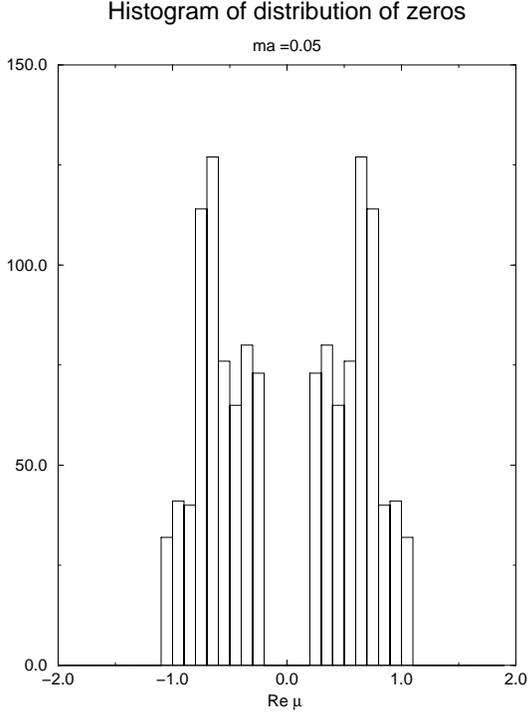}
    \caption[xxx]{ Histogram of the real parts of the 
zeros in the complex $\mu$-plane at strong coupling and bare
quark mass, $ma=0.05$ on a $6^4$ lattice.}

\label{fig:zerm0.05}
\end{figure}

\begin{figure}[htb]
\epsfxsize=7.0cm
\epsfbox{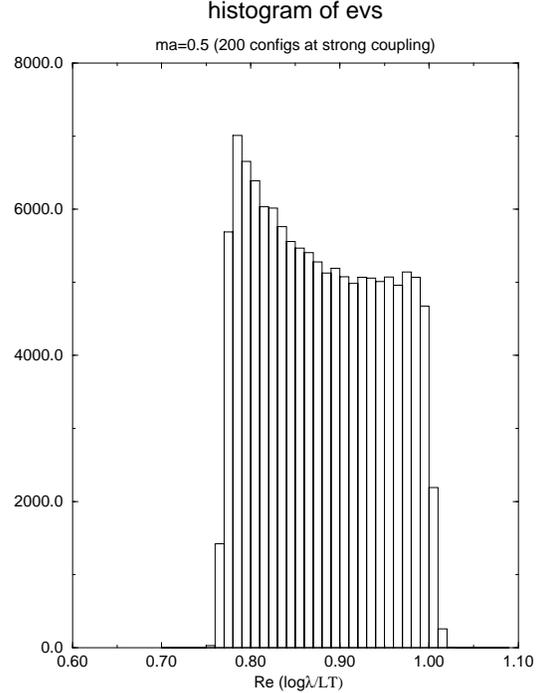}
    \caption[xxx]{ Histogram of the real parts of the eigenvalues 
in the complex $\mu$-plane at strong coupling and bare
quark mass, $ma=0.5$ on a $6^4$ lattice.}
\label{fig:eigm0.5}

\end{figure}

\begin{figure}[htb]
\epsfxsize=7.0cm
\epsfbox{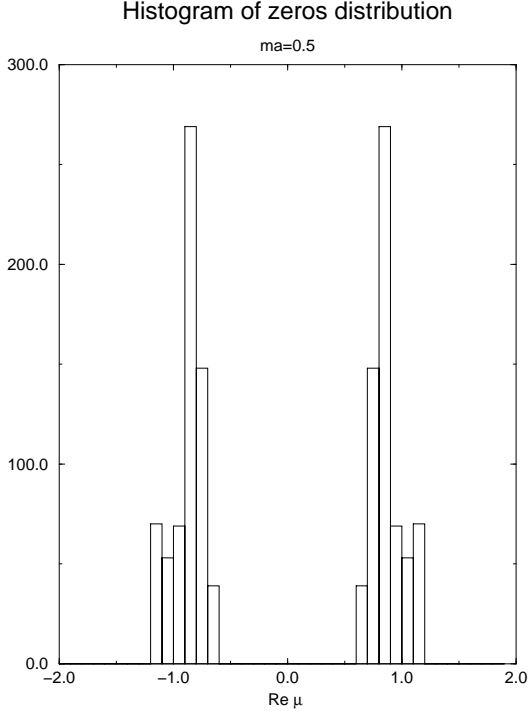}
    \caption[xxx]{ Histogram of the real parts 
of the zeros in the complex $\mu$-plane at strong coupling and bare
quark mass, $ma=0.5$ on a $6^4$ lattice.}

    \label{fig:zerm0.5}
\end{figure}

This difference in distribution between the eigenvalues and the zeros is found at
all other bare quark masses. Fig. \ref{fig:eigm0.05} and Fig. \ref{fig:zerm0.05}   show
our results at quark mass $m=0.05$ while Fig. \ref{fig:eigm0.5} and 
Fig. \ref{fig:zerm0.5} show results at $m=0.5$.

\vspace{0.2cm}

\vspace{0.2cm}

\begin{figure}[htb]
\epsfxsize=7.0cm
\epsfbox{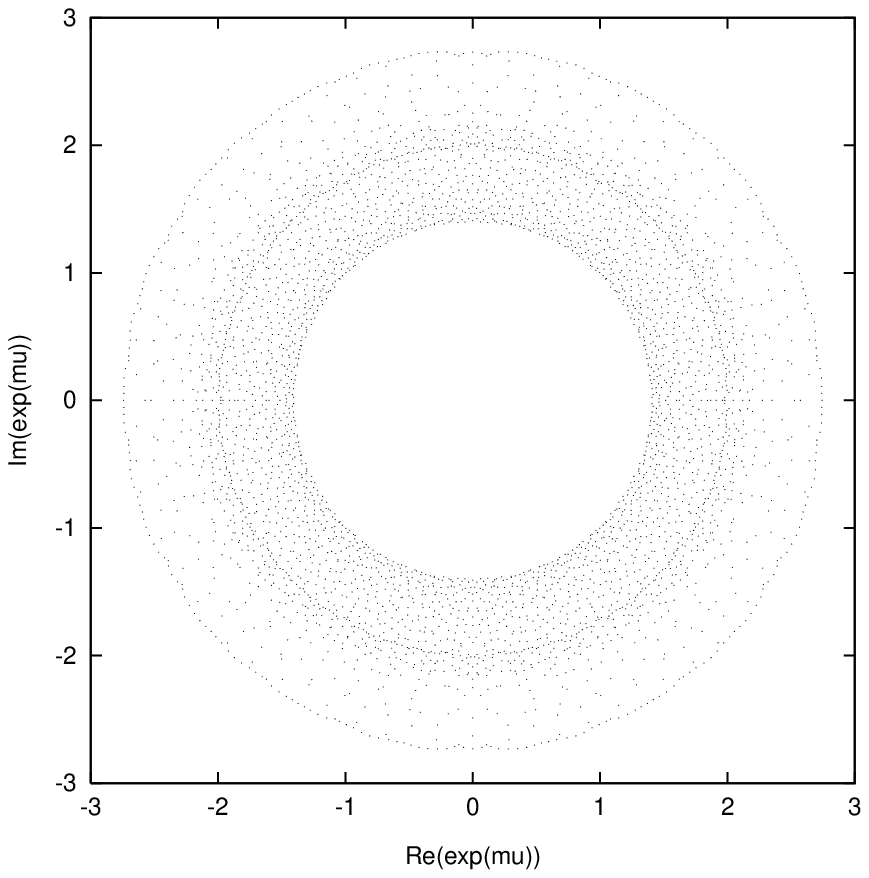}
\caption[xxx]{Zeros in the $e^\mu$ plane
for $m = .095$ , $6^4$ lattice . The critical line is
the thin line inside the denser region $e^{\mu} = e^{\mu_c}$}
\label{fig:zero6}
\end{figure}

\begin{figure}[htb]
\epsfxsize=10.0cm
\epsfbox{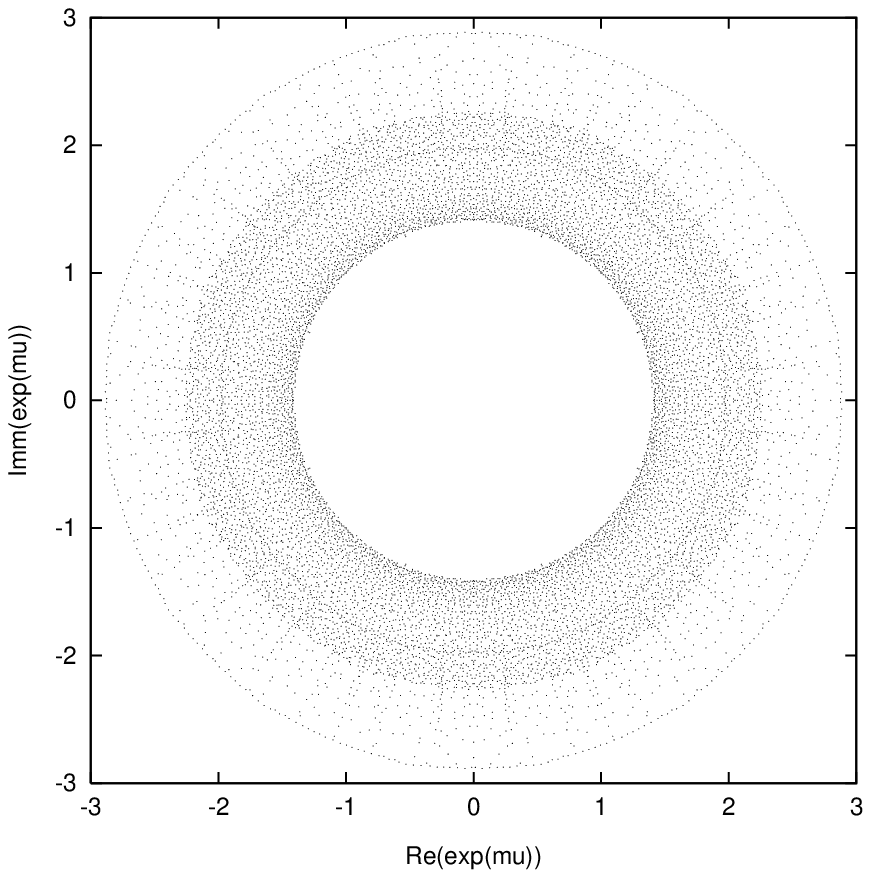}
    \caption[xxx]{%
Zeros in the $e^\mu$ plane
for  $m = .1$  $8^4$ lattice.}
    \label{fig:zero8}
\end{figure}

Figs. \ref{fig:zero6} and \ref{fig:zero8}
show clearly the banded structure of the zeros distribution in the 
$e^{\mu}$ plane for two different lattice sizes and quark masses.


\begin{figure}[htb]
\epsfxsize=7.0cm
\epsfbox{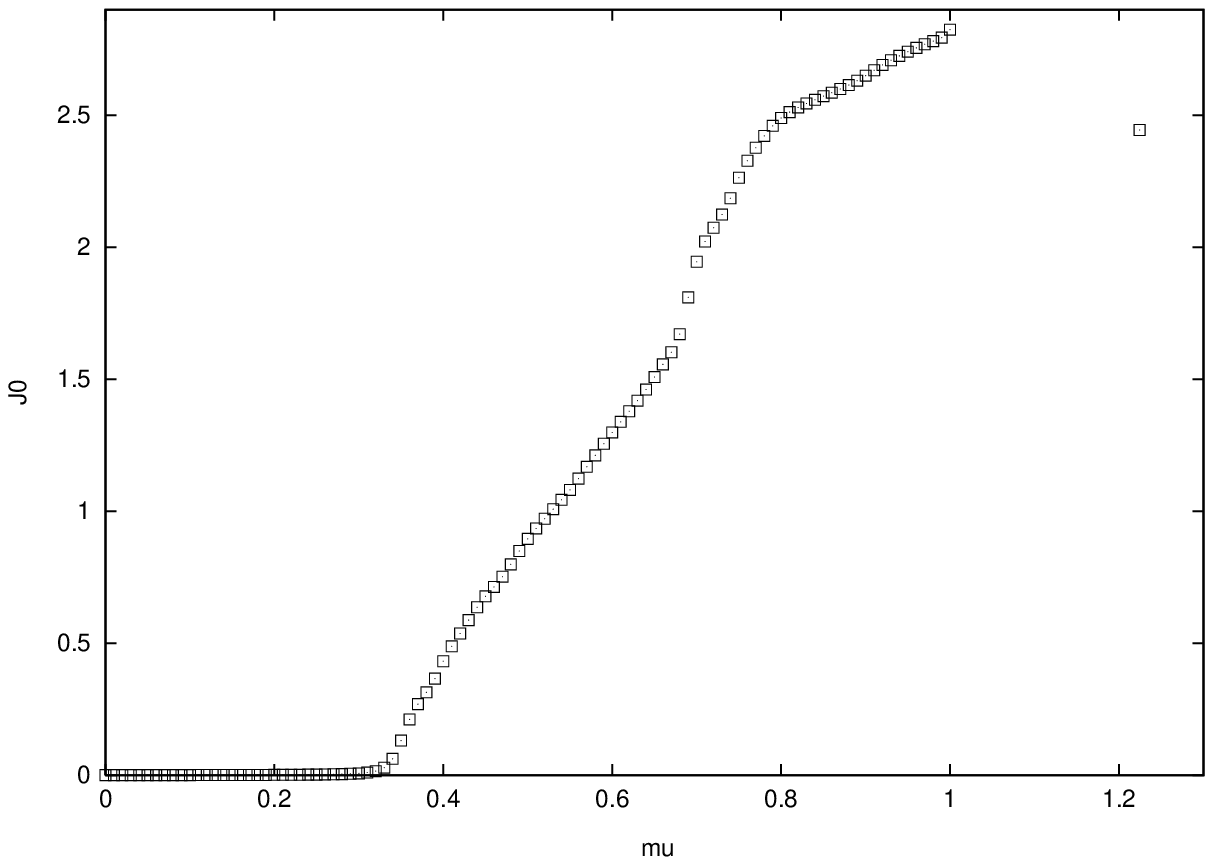}
\epsfxsize=7.0cm
\epsfbox{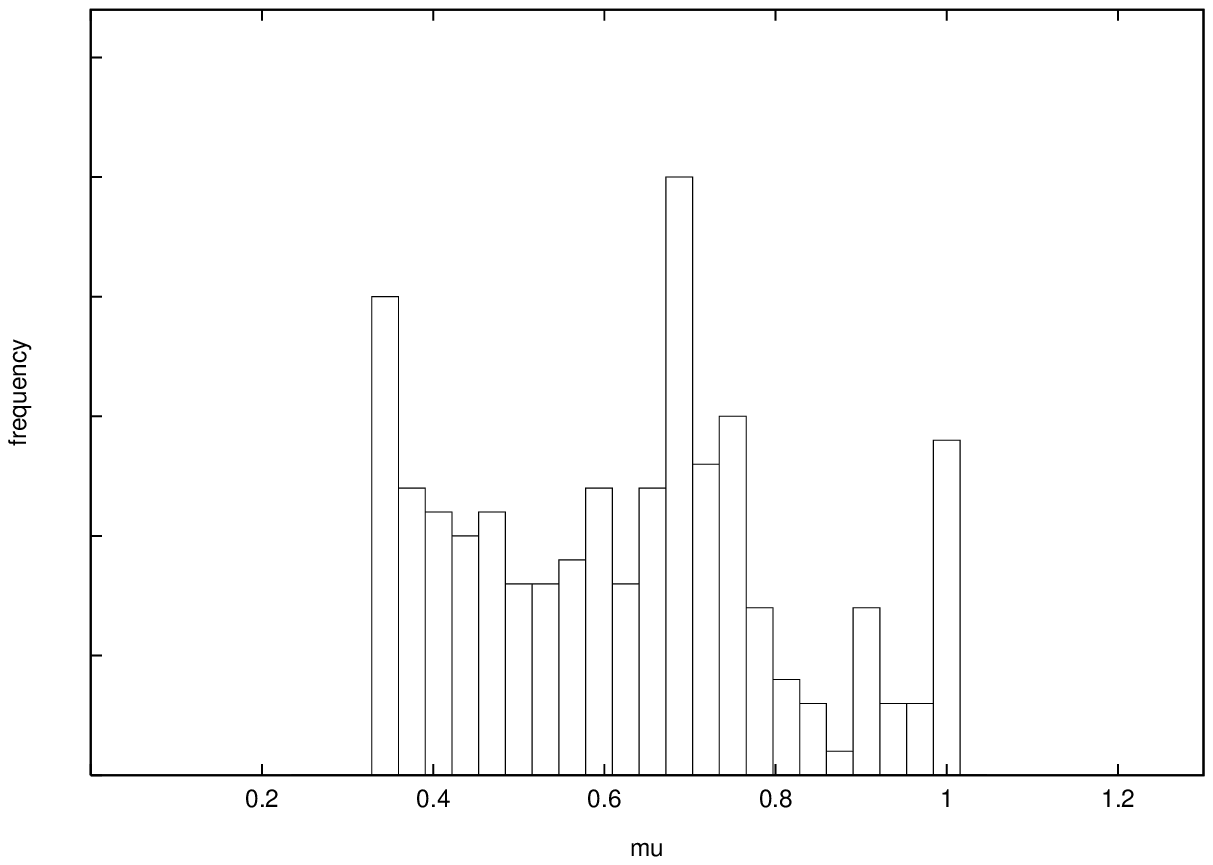}
    \caption[xxx]{%
      The number density at quark
mass $m_q = .1$  (upper) on a $6^4$ lattice at strong coupling.
The behaviour of the number density is reflected in the
histogram of the real parts of the zeros in the complex $\mu$-plane
shown in the lower portion of the figure.}
    \label{fig:overview}
\end{figure}

Fig. \ref{fig:overview} compares the distribution of the real part of the zeros
with the number density (on a $6^4$ lattice with $m=0.1$). Again the
histogram for the zeros shows three distinct peaks:
$\mu_{o}\simeq 0.3$ corresponding to the onset of net non-zero quark
density; $\mu_{c}\simeq 0.7$ corresponding to the small discontinuity in the number density and
the expected critical chemical potential; $\mu_{s}\simeq 1.0$ corresponding to the
lattice saturation point. Comparison of the two plots in this figure
shows that the derivative of the number density correlates well with 
the density of states (frequency histogram). 

\begin{figure}[htb]
\epsfxsize=5.7cm
\rotate[r]
{\epsfbox{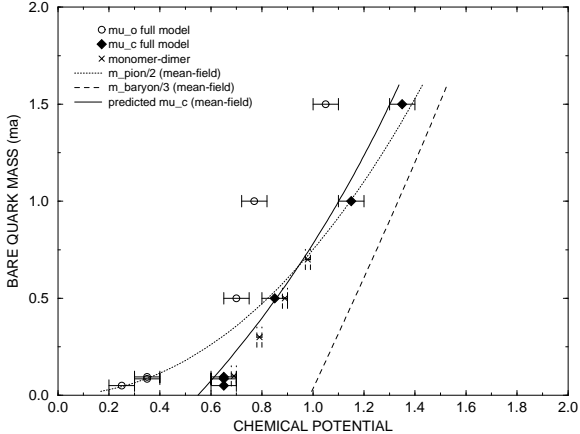}}
\caption[xxx]{Summary of the
results for the critical point $\mu_c$, and
current onset $\mu_o$. $\mu_c$ follows the prediction of the mean
field analysis of ref.[4]. 
The onset is close to half the pion mass at small mass, 
and  lower for $m_q > .5$ . The data for the pion mass are from ref.
[7]}
    \label{fig:onsets_b}
\end{figure}

Fig. \ref{fig:onsets_b}
gives an overview of the location of $\mu_o$ and $\mu_c$ for six quark
masses in the range $ma=0.05$ to $1.5$. The scaling of the central peak
of the histogram is in remarkable agreement with the mean-field
prediction calculated form eqn.13 and is consistent with the critical
$\mu$ predicted by the monomer-dimer simulations. The peak corresponding
to $\mu_o$ is consistent with the mean-field pion threshold for the smaller
quark masses but significantly lower for $ma=1.0,1.5$.

\section{Lattice QCD at Intermediate Coupling}

\subsection{$Z(3)$ Tunnelling}
\begin{figure}[htb]
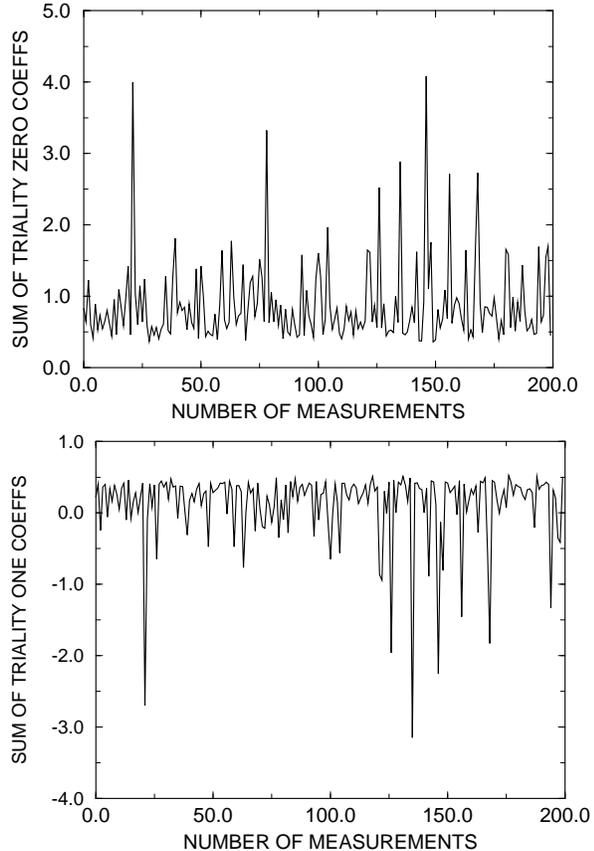

\vspace{-0.6cm}
\epsfxsize=5.5cm
\rotate[r]
{\epsfbox{8tun0.epsi}}
\vspace{0.2cm}
\epsfxsize=5.5cm
\rotate[r]
{\epsfbox{8tun1.epsi}}
\vspace{0.2cm}
\vspace{-0.6cm}
\caption{The HMC time evolution of the expansion coefficients on an $8^4$ lattice 
at $\beta=5.1$ and $ma=0.01$. The behaviour of the triality 2 coefficients is similar to that
of the triality 1 coefficients.}
\vspace{-0.6cm}
\label{fig:tunn}
 \end{figure}

In the pure gauge theory we have $N_c(=3)$ equivalent vacua related by $Z(N_c)$
rotations.
 Tunnelling between the different $Z(3)$ vacua is much more probable
in the confined sector than in the deconfined sector. Since the pure-gauge action as well
as
 the integration measure are invariant under the $Z(3)$
transformation, the GCPF can also be written as:
\begin{equation}
{\cZ}(\mu)={{\int [dU][dU^\dagger]det M(\mu+z_3/n_t)e^{-S_g[U,U^\dagger]}}\over {
\int [dU][dU^\dagger]det M(\mu=0)e^{-S_g[U,U^\dagger]}}}
\end{equation}
Averaging over the three $Z(3)$ vacua
would eliminate the triality non-zero coefficients and, since $Z(0)=1$, the sum of
the
 triality zero
coefficients should tend to 1.
As Fig. \ref{fig:tunn} shows,at $\beta=5.1$ and $m=0.01$, we obtained a clear signal for Z(3) tunnelling. 
The sum of the triality zero coefficients shows a strong tendency to 
average to 1 whilst the sum of the triality
one (and two) coefficients tends to zero. This  is consistent with
simulating in the confined phase at $\mu=0$ as expected at this coupling and quark mass.

\subsection{Simulations on  $6^4$ and $8^4$ lattices at intermediate coupling}

We have investigated\cite{SBK} the chemical potential (the onset $\mu$)
required to make the level with 3
quarks equally probable with that of the zero quark level, i.e. $\mu_{onset}=(\epsilon_3-\epsilon_0)/3.$
This ad-hoc definition takes into account that it is these energy levels which are most
accurately determined and allows errors to be estimated directly.

\begin{figure}[htb]
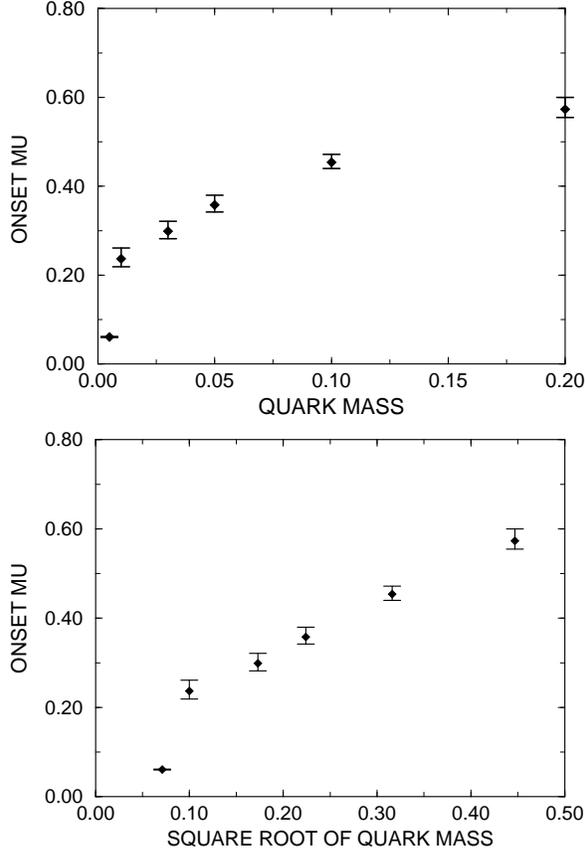

\epsfxsize=5.5cm
\rotate[r]
{\epsfbox{qmass.epsi}}
\vspace{0.2cm}
\epsfxsize=5.5cm
\rotate[r]
{\epsfbox{sqrtqmass.epsi}}
\vspace{-0.6cm}
\caption{The quark mass dependence of the onset $\mu$ ($\beta=5.1$ on a $6^4$ lattice). }
\vspace{-0.6cm}
\label{fig:qmass}
 \end{figure}

Fig. \ref{fig:qmass} shows the dependence of
the onset $\mu$ on the quark mass and on the square root of the 
quark mass on a $6^4$ lattice at gauge
coupling $\beta=5.1$.
As at strong coupling, there is a strong signal that this onset $\mu$ is
dependent on the mass of a Goldstone
boson for quark mass $m > 0.01$. (On a $6^4$ lattice, at this $\beta$, the system
becomes 'deconfined' at some $m < 0.01$. We have checked this behaviour by determining its Lee-Yang
zeros in the complex $m$-plane and they are complex with non-zero real part $ <  0.01$.)

\begin{figure}[htb]
\centering
\leavevmode
\epsfxsize=5.5cm
\rotate[r]
{\epsfbox{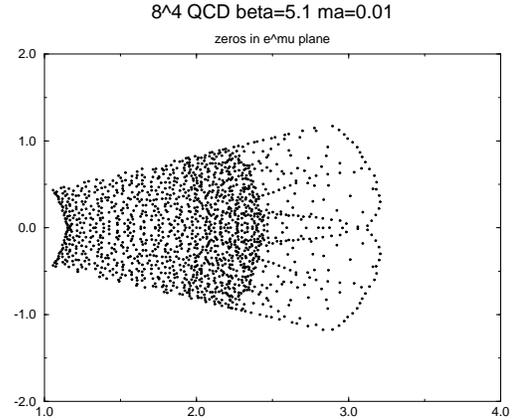}}
\caption{The Lee-Yang zeros on a $8^4$ lattice in the complex $e^{\mu}$ plane at $\beta=5.1$
and bare quark mass 0.01. Only one segment is shown but the intrinsic $Z(n_t=8)$ of Figs.
4 and 5. also applies here.}
\label{fig:zeros8}
\end{figure}

Fig. \ref{fig:zeros8} shows the distribution of the zeros found on a $8^4$ lattice at the same $\beta$ and
quark mass $0.01$ in the $e^{\mu}$ plane.  The zeros associated with the onset $\mu$ lie on
the arc of a circle (this circular pattern is also observed in the strong
coupling results described above). Again the distribution is consistent with a
number density becoming non-zero at a small $\mu$. Also we see a circular band of
zeros developing at $0.7 < \mu < 0.9$. We believe that this is a signal for the
physical $\mu_c$.

\section{The U(1) Gross-Neveu Model }

\subsection{The Lattice Formulation}

In the above simulations of QCD, the initial onset of the critical behaviour 
at non-zero $\mu$ was controlled by the pion mass 
rather than that of the lightest baryon. To check if this is an artifact arising 
from measuring our observables on an ensemble generated at $\mu=0$ we have 
investigated the above methods in a study of the
Gross-Neveu model at finite density. The lattice formulation of this model is described in
ref.\cite{HandsKK95}

The lattice action for the bosonized Gross-Neveu model with U(1) chiral symmetry is
\begin{eqnarray}
S & = &\sum_{i=1}^{N_f/4}\biggl[\sum_{x,y}\bar\chi_i(x){\cal M}_{x,y}\chi_i(y) \nonumber \\
  & + & 
{1\over8}\sum_x\bar\chi_i(x)\chi_i(x)
\Bigl(\sum_{<\tilde x,x>}\sigma(\tilde x) \nonumber \\
& + & i\varepsilon(x)
\sum_{<\tilde x,x>}
\pi(\tilde x)\Bigr)\biggr] \nonumber \\
  & + &{N_f\over8g^2}\sum_{\tilde x}
(\sigma^2(\tilde x)+\pi^2(\tilde x)).
\end{eqnarray}
Here, $\chi_i$ and $\bar\chi_i$ are complex Grassmann-valued staggered
fermion fields defined on the lattice sites, the auxiliary scalar and
pseudoscalar fields $\sigma$ and $\pi$ are defined on the dual lattice
sites, and the symbol $<\tilde x,x>$ denotes the set of 8 dual sites
$\tilde x$ adjacent to the direct lattice site $x$. $N_f$ is the number of
physical fermion species.

The fermion kinetic operator ${\cal M}$ is given by
\begin{eqnarray}
{\cal M}_{x,y} & = & {1\over2}\Bigl[\delta_{y,x+\hat0}e^\mu-
\delta_{y,x-\hat0}e^{-\mu}\Bigr] \nonumber \\
& + & {1\over2}\sum_{\nu=1,2}\eta_\nu(x)
\Bigl[\delta_{y,x+\hat\nu}-\delta_{y,x-\hat\nu}\Bigr] \nonumber \\
& + & m\delta_{y,x},
\end{eqnarray}
where $m$ is the bare fermion mass, $\mu$ is the chemical potential, and
$\eta_\nu(x)$ are the Kawamoto-Smit phases
$(-1)^{x_0+\cdots+x_{\nu-1}}$. The symbol $\varepsilon(x)$ denotes the
alternating phase $(-1)^{x_0+x_1+x_2}$.

This lattice model (at non-zero lattice spacing) has the symmetry:
${\rm U}(N_f/4)_V\otimes{\rm U}(N_f/4)_V\otimes{\rm U}(1)_A$.
It is the ${\rm U}(1)_A$ symmetry which is
broken, either spontaneously by the dynamics of the system, or
explicitly by a bare fermion mass. We simulated the $N_f=12$ model
corresponding to three lattice species.

The expansion of this GCPF is very similar to that described above for
standard QCD. At finite density the Dirac fermion matrices $M$ and $\hat{M}$ are given by:
\begin{eqnarray}
       2iM_{xy}(\mu)&=&Y_{xy} + G_{xy} + V_{xy} e^{\mu} +V^{\dag}_{xy} e^{-\mu} \nonumber \\
-2i\hat{M}_{xy}(\mu)&=&Y_{xy}^{\dag} + G_{xy} + V_{xy} e^{\mu} +V_{xy}^{\dag} e^{-\mu}. \nonumber
\end{eqnarray}
where $Y = 2i( m_q + \frac{1}{8} \sum_{<x,\tilde x>} (\sigma( \tilde x)
                   +i\epsilon\pi( \tilde x)))\delta_{xy}. \nonumber$

The determinants of these fermion matrices are related to that of
the propagator matrix
\begin{equation}
P=\left(\begin{array}{cc}
-GV-YV & V \\
  -V   & 0
\end{array} \right)
\end{equation}
by
$\det(2iM) = e^{3 \mu n_s^3 n_t} \det(P-e^{-\mu})$ and
$\det(2i\hat{M})=e^{3 \mu n_s^3 n_t} \det( (P^{-1})^{\dag} - e^{-\mu})$.

As for standard QCD, determination of the eigenvalues of $P^{n_t}$ gives the expansion
for the GCPF:
\begin{eqnarray}
{Z}& = & \sum_{n=-2 {n_s}^2 }^{2 {n_s}^2}<b_{|n|}>e^{n\mu {n_t}}  \nonumber \\
 & = & \sum_{n=-2 {n_s}^2}^{2 {n_s}^2}e^{-(\epsilon_n - n\mu)/T}.
\end{eqnarray}

\subsection{Results}
This model has a massless pion and its fermion determinant is real and
non-negative even when $\mu \neq 0$. Therefore standard Monte-Carlo algorithms can
be used to study its critical behaviour as a function of $\mu$.

In particular,
at an inverse four-fermion coupling of 0.5 and bare fermion mass $m=0.01$,
the theory has a critical chemical potential $\mu_c=0.725(25)$ at which the chiral
symmetry is restored. Note, the Gross-Neveu model is non-confining and the gauge fields
do not enter into the dynamics. 

\begin{figure}[htb]
\centering
\leavevmode
\epsfxsize=5.5cm
\rotate[r]
{\epsfbox{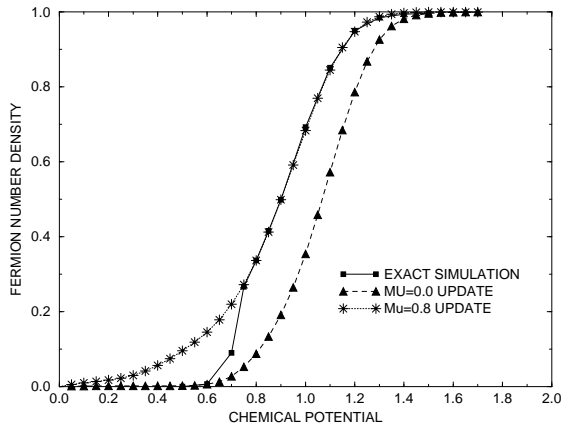}}
\caption{The fermion number density vs. the chemical potential, $\mu$, on a $16^3$
lattice at an inverse four-fermion coupling 0.5 and fermion mass 0.01}
\label{G-N_number_density}
 \end{figure}

\begin{figure}[htb]
\centering
\leavevmode
\epsfxsize=5.5cm
\rotate[r]
{\epsfbox{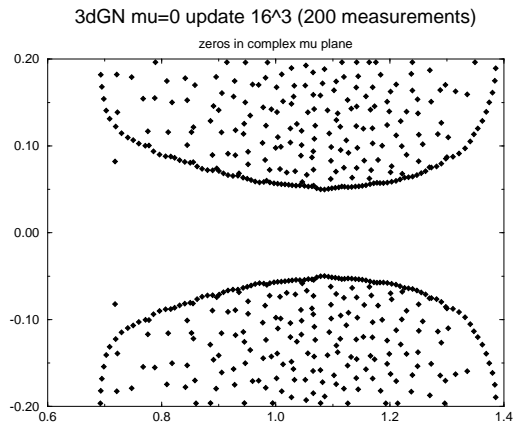}}
\caption{ The Lee-Yang zeros in the complex $\mu$ plane for the Gross-Neveu model at an
inverse four-fermion coupling 0.5 and fermion mass 0.01 on a $16^3$ lattice}
\label{fig:3dGNmu0.0zeros}
 \end{figure}

We performed simulations developing the ensemble at $\mu=0$ to confirm that Eqn. 5
does give an extrapolation which predicts the correct critical behaviour.
The number density for a $16^3$ lattice at the above coupling and fermion mass is shown in
Fig. \ref{G-N_number_density}. The chemical potential at which the number density, $<n>$, becomes non
-zero, $\mu_o$, is
correctly predicted. Comparison with the exact simulation\cite{HandsKK95}
(in which generation of the ensemble was at the $\mu$ at which the measurement was made)
shows that at $\mu > \mu_o$
the number density is underestimated and the discontinuity in $<n>$ is absent.
The Lee-Yang zeros in the complex $\mu$ plane are the zeros of Eqn. 5 and should signal
the critical $\mu$ associated with the chiral transition transition.
They are shown in Fig. \ref{fig:3dGNmu0.0zeros} There is a line formed by 6 zeros, two of which are
isolated from the rest of the distribution. The line intersects the real $\mu$ axis at
$\mu=0.72$, in agreement with the critical $\mu$ found by the exact method.

Hence we find that the onset $\mu$ found via the number density is
not controlled by the Goldstone boson, 
and the critical $\mu$ is given by the distribution of the Lee-Yang zeros.

Fig. \ref{G-N_number_density}
also shows the number density obtained from an ensemble generated at an update
$\mu=0.8$, greater than $\mu_c$. In the chiral limit it is consistent with that of a free gas of 
massless fermions whereas the number density from the $\mu=0$ update is that of a free
gas of massive fermions. The exact results show the transition
between these two phases at $\mu_c$.

\section{Conclusions}
Simulation of standard lattice QCD at finite density seems to be plagued by
the problems of the naive quenched version of the theory, namely the chemical
potential at which the fermion number begins to differ from zero is
controlled by a Goldstone pion rather than by the lightest baryon.

Clearly questions must be answered:

Why do we continue to observe an onset chemical potential which scales with
the pion mass? Is the band of zeros at higher $\mu$ a signal of the
transition in the number density? Is this band associated with the deconfinement
and chiral transitions? Do these two transitions occur at the same $\mu$?

The success of the monomer dimer approach (of use only at strong coupling) may
be due to analytic integration first over the gauge fields, and then the fermion
fields, giving an effective action with a reduced sign problem. Standard Hybrid
Monte-Carlo algorithms integrate the fermion fields prior to the gauge fields.
 We are now attempting to
understand why at strong coupling, we find an onset chemical potential, $\mu_o$, in addition
to the expected signal at $\mu_c$.

Incomplete cancellation could arise from either
limited statistics or from the reweighting procedure itself. Using an
ensemble created at $\mu=0$, together with reweighting with respect to the fermion
determinant at $\mu=0$ may not give sufficient overlap with an ensemble generated 
at $\mu \neq 0$ to correctly describe the true physics there. Although
a limited increase in statistics revealed that the peak in the histogram
of zeros corresponding to $\mu_o$ grew proportionally with the peak at
$\mu_c$ it is still possible that the signal for $\mu_o$ will be
cancelled out in the limit of infinite statistics {\it once the $Z(3)$
invariance has been completely achieved}.

Another solution may be to approach QCD by adding a perturbatively irrelevant 
four-fermion interaction term
to the action and to study the critical nature of that theory as this additional term
is weakened. This approach permits simulations directly in the chiral limit.

\section{Acknowledgements}
        We would like to thank Misha Stephanov
for very useful suggestions. This work was supported
in part by the NSF through grant NSF-PHY92-00148,by DOE, by
PPARC through grant GR/K55554 and by NATO via grant CRG960002.
Computations were performed on the C90s at NERSC and PSC.

\end{document}